\documentclass[12pt]{article}

\usepackage{amssymb}
\usepackage{amsmath}
\usepackage{epsfig}

\def\be{\begin{equation}}
\def\ee{\end{equation}}

\begin{document}
\begin{titlepage}
\thispagestyle{empty}
\hskip 1 cm
\vskip 0.5cm

\vspace{25pt}
\begin{center}
    { \LARGE{\bf Exact Friedmann Solutions in
    Higher-Order Gravity Theories}}
    \vspace{33pt}

  {\large  {\bf   Timothy Clifton\footnote{T.Clifton@cantab.net}}}

    \vspace{15pt}

 {Department of Physics, Stanford University, Stanford, CA 94305, USA.}

  \end{center}

   \vspace{20pt}

\begin{abstract}
We find the general behaviour of homogeneous and isotropic
cosmological models in some fourth-order theories of gravity.
Explicit, exact, general solutions are given for both empty universes
and those filled with a perfect fluid.  For the vacuum case, solutions
are found with closed, open and flat geometries, whilst the perfect fluid
solutions are all spatially flat.  Both early and late-time limits are
studied, and attractor behaviour towards simple power-law expansion is
identified.  Multiple solutions to the same theories, with the same
matter content and topology are found.  It is shown that these
solutions exhibit great variety in their evolution.
\end{abstract}

\vspace{10pt}
\end{titlepage}

\tableofcontents

\newpage

\section{Introduction}

Generalisations of the usual Einstein-Hilbert action of general
relativity (GR) have been extensively studied in the literature.  One
frequently considered modification is to replace the Ricci scalar, $R$, by
some analytic function, $f(R)$ (see e.g \cite{buch, kerner, BO,
  magg}).  Theories of this type are referred to as fourth-order,
as the field equations generated from them are generically
fourth-order in derivatives of the metric.  Motivations for such
studies are found from various different sources.  One often cited
reason is that early attempts to create a perturbatively
renormalizable quantum field theory of gravity found success by adding
extra terms, quadratic in the Ricci curvature, to the action
\cite{pech}.  More recently, the effective actions of some string theories
have been shown to include higher-order curvature terms
\cite{str1,str2}.  Studies of $f(R)$ theories have also been
performed in cosmological settings, often in attempts to better
understand the late-time accelerating expansion of the universe \cite{tur, new},
cosmological inflation \cite{berk, brun, herv} or the nature of an initial singularity
\cite{brrw, clftn, dnsby}.

Difficulty can arise, in studies of fourth-order theories, as the
field equations involved are considerably more complex than their counterparts
in GR.  This extra complexity brings new and interesting behaviour,
such as violations of Birkhoff's theorem \cite{birk} or violations of
the no hair theorems of de Sitter space, \cite{herv}.  However, this
extra complexity also makes it even more difficult to find exact
solutions of the field equations.  Whilst some exact solutions have been
found in the spherically symmetric \cite{birk, power} and cosmological
\cite{power, godel, Buc70, schmidt, Car04, CB} situations, these are
all particular solutions.  To date, there have been no general
solutions published in the literature, for any set of non-maximal
symmetries.  Exact solutions, and particularly exact general
solutions, are of great importance for understanding a theory.  In
this paper we will present homogeneous and isotropic exact general solutions, obtained
through direct integration of the field equations, to some
fourth-order theories of gravity.  We expect these solutions to be of
use for helping to understand fourth-order theories of gravity, and the
evolution of the universes they govern.

\begin{figure}[ht]
\center \epsfig{file=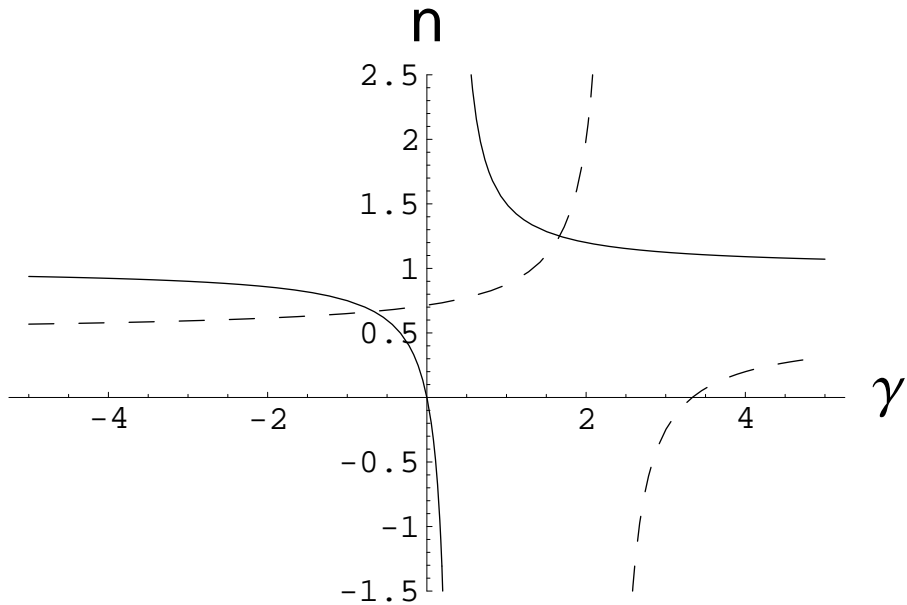,height=8cm}
\caption{A plot of $n=3 \gamma/(3
\gamma-1)$ (solid line) and $n = (10-3\gamma)/(2 (7-3 \gamma))$
(dashed line).  For these values of $n$ the field equations for
homogeneous and isotropic perfect fluid cosmologies can be integrated
directly.}
\label{deltaplot}
\end{figure}

Investigations of fourth-order theories usually follow one of
two approaches: Either they take the full-theory and look for
approximate solutions, or they approximate the
theory and look for exact solutions.  We will take the later approach
and consider theories that have a Lagrangian proportional to
$R^n$.  Such theories are scale invariant, and reduce to GR in the
limit $n\rightarrow 1$.  These theories may be considered as the limit
of a more general Lagrangian that has a power of $R$ dominating in
some particular regime, or as a simple deviation from the standard theory in
their own right.  We find that for homogeneous and isotropic
cosmologies the field equations of these theories can often be
integrated directly.  For spatially flat vacuum cosmologies the
general solution can be found for any $n$, for spatially curved
vacuum cosmologies we find the general solutions when $n=3/2$, and for
spatially flat perfect fluid cosmologies (with equation of state
$p=(\gamma-1)\rho$) we find the general solutions when $n=3 \gamma/(3
\gamma-1)$  and when $n = (10-3\gamma)/(2 (7-3 \gamma))$.  These
dependences are shown graphically in figure \ref{deltaplot}, below.

Investigations of homogeneous and isotropic cosmologies in $R^n$
theories have been performed before in \cite{power, Car04}.  In
these papers the authors find power-law exact, particular solutions
and perform a dynamical systems analysis of the phase space of more
general solutions.  These studies show that the general solutions are
often attracted to simple power-law solutions, at both late and
early times.  We will confirm this behaviour here, and add to the
previous work by finding explicit expressions for the general
evolution of these universes.  

We will now comment on the extent to which the solutions found in this
paper can be considered candidates for the description of our universe.  It has been shown in
\cite{birk} and \cite{power} that gravitational theories derived from a Lagrangian of
the form $R^n$ are constrained by observations in our local universe to have $n$ very close
to $1$.  It has also been suggested (see e.g. \cite{Od1},
\cite{Od2}) that it may not be possible for general $f(R)$ theories of
gravity to transition between eras dominated by different fluids in
the same way that occurs in the usual general relativistic
description.  These results suggest that our neighbourhood of
space-time should be well described by a theory that is very close to the
Einstein-Hilbert one.  However, with these constraints in mind, there
are still good reasons to be interested in deviations of the $R^n$
form.  Firstly, even if $n$ is very tightly constrained by local
observations, any small deviation from $n=1$ may produce
significant deviations at early times in the universe's history.
This will be shown explicitly below.  Secondly,
theories of the type $R+\alpha R^n$ are often considered in the
literature.  If $n> 1$ then we may expect the effective gravitational theory
in the early universe to be well described by a Lagrangian of
the form $R^n$.  In such cases a more appropriate description of the
evolution of the early universe would be given by an $R^n$ theory, even though
observations from our local universe show strong agreement
with an effective theory of the Einstein-Hilbert type.

This paper will proceed as follows.  In section 2 we will give the
field equations for the theory, and show how they can be simplified by
recasting them in terms of new variables and transforming time
coordinates.  Simple power-law exact, particular solutions to these
equations are given. In section 3 we show how the field equations for vacuum
cosmologies can be decoupled by a further transformation of
variables.  These equations are then integrated directly, and the
solutions given explicitly in terms of the metric.  Section 4 follows
a similar prescription, this time solving the field equations in the
presence of a perfect fluid.  In section 5 we perform a brief
analysis of the vacuum solutions, and in section 6 we analyse the
perfect fluid solutions.  Section 7 provides a closing discussion.
For readers interested in the solutions, but not the
derivations, the solutions in sections 3 and 4 are boxed to make them
easily identifiable.

\section{Field equations}

We consider here a gravitational theory derived from the Lagrangian density 
\begin{equation}
\mathcal{L}_{G}=\frac{1}{\chi }\sqrt{-g}R^{1+\delta },  \label{action}
\end{equation}
where $\delta $ is a real number and $\chi $ is a constant. The limit
$\delta \rightarrow 0$ gives the Einstein-Hilbert
Lagrangian of GR, and we are interested in isotropic
and homogeneous cosmological solutions with $\delta \neq 0$.  Theories
of this kind are dynamically equivalent to scalar-tensor theories (see
e.g. \cite{ST}).  

We denote the matter action as $S_{m}$ and ignore the boundary term.  Extremizing 
\begin{equation*}
S=\int \mathcal{L}_{G}d^{4}x+S_{m},
\end{equation*}%
with respect to the metric $g_{ab}$, then gives \cite{Buc70} 
\begin{multline}
\delta (1-\delta ^{2})R^{\delta }\frac{R_{,a}R_{,b}}{R^{2}}-\delta (1+\delta
)R^{\delta }\frac{R_{;ab}}{R}+(1+\delta )R^{\delta }R_{ab}-\frac{1}{2}%
g_{ab}RR^{\delta }  \label{field} \\
-g_{ab}\delta (1-\delta ^{2})R^{\delta }\frac{R_{,c}R_{,}^{\ c}}{R^{2}}%
+\delta (1+\delta )g_{ab}R^{\delta }\frac{\Box R}{R}=\frac{\chi }{2}T_{ab},
\end{multline}%
where $T_{ab}$ is the energy--momentum tensor of the matter fields, and is defined
in terms of $S_{m}$ and $g_{ab}$ in the usual way. We take the
quantity $R^{\delta }$ to be the positive real root of $\vert R \vert$.

We are concerned with idealised homogeneous and
isotropic space-times described by the Friedmann-Robertson-Walker metric
with spatial curvature parameter $\kappa $: 
\begin{equation}
ds^{2}=-dt^{2}+a^{2}(t)\left( \frac{dr^{2}}{(1-\kappa r^{2})}+r^{2}d\theta
^{2}+r^{2}\sin ^{2}\theta d\phi ^{2}\right) .  \label{FRW}
\end{equation}%
Substituting this metric ansatz into the field equations (\ref{field}), and
assuming the universe to be filled with a perfect fluid of pressure $p$ and
density $\rho $, gives the generalised version of the Friedmann equations 
\begin{align}
(1-\delta )R^{1+\delta }+3\delta (1+\delta )R^{\delta }\left( \frac{\ddot{R}
}{R}+3\frac{\dot{a}}{a}\frac{\dot{R}}{R}\right) -3\delta (1-\delta
^{2})R^{\delta }\frac{\dot{R}^{2}}{R^{2}}& =\frac{\chi }{2}(\rho -3p)
\label{Friedman1} \\
-3\frac{\ddot{a}}{a}(1+\delta )R^{\delta }+\frac{R^{1+\delta }}{2}+3\delta
(1+\delta )\frac{\dot{a}}{a}\frac{\dot{R}}{R}R^{\delta }& =\frac{\chi }{2}
\rho
\label{Friedman2}
\end{align}
where, as usual, 
\begin{equation}
R=6\frac{\ddot{a}}{a}+6\frac{\dot{a}^{2}}{a^{2}}+6\frac{\kappa }{a^{2}}.
\label{R2}
\end{equation}
It can be seen that in the limit $\delta \rightarrow 0$ these equations
reduce to the standard Friedmann equations of GR. A study of
the vacuum solutions to these equations has been made by
Schmidt, see the review \cite{schmidt}, and a study of the
perfect-fluid evolution has been made by Carloni et al 
\cite{Car04}.  In \cite{power} Clifton and Barrow investigated
solutions when $\kappa=0$, and showed that attractor power-law
solutions exist.  Various conclusions are also immediate from the general
analysis of $f(R)$ Lagrangians made in \cite{BO} by specialising them to
the case $f=R^{1+\delta }$.  These previous studies have focussed on
the existence of particular exact solutions, or qualitative
investigations of the phase space of general solutions.  In what
follows we shall be interested in determining the general evolution of
$a$ in terms of explicit, exact solutions.  This should lead to fresh
insights into the evolution of universes described by these theories.

Assuming a perfect-fluid equation of state of the form $p=(\gamma-1)
\rho $, the usual conservation equation gives $\rho = \rho_0 a^{-3\gamma}$.
Substituting this into equations (\ref{Friedman1}) and (\ref{Friedman2}),
with $\kappa =0$, gives the particular power-law exact Friedmann solution, for $\gamma
\neq 0$,
\begin{equation}
a(t)=t^{\frac{2(1+\delta )}{3\gamma}}.  \label{power}
\end{equation}
Alternatively, if $\gamma =0$, there exists the de Sitter solution
$a(t)=e^{m t}$.  These power-law solutions reduce to the usual general
relativistic solutions in the limit $\delta \rightarrow 0$.
Furthermore, the power-law solutions
\begin{equation}
a(t)= t^{\frac{\delta (1+2\delta)}{(1-\delta)}} \label{power2}
\end{equation}
and
\begin{equation}
a(t) = t^{\frac{1}{2}} \label{power3}
\end{equation}
also exist.  These solutions are independent of $\gamma$, and
therefore of the matter content of the universe.  The $\delta =0$
limit of (\ref{power2}) clearly corresponds to Minkowski space.  These
three particular solutions were both shown in \cite{power} to
be attractors of the spatially flat general solution, under certain
conditions.  In what follows we will be able to show this attractor
explicitly in terms of exact general solutions.

The equations (\ref{Friedman1}), (\ref{Friedman2}) and (\ref{R2}) can
be cast in a simpler form by defining the new variables 
\be
\nonumber
\phi \equiv \sqrt{3} \ln R^{\delta} \qquad \qquad \text{and} \qquad
\qquad \bar{a}\equiv e^{\phi /2 \sqrt{3}} a.
\ee
Transforming to the time coordinate 
\be
\nonumber
d \eta \equiv e^{\phi /2 \sqrt{3}} dt 
\ee
and setting $\chi =1$ we then have two
second-order evolution equations, for the two variables $\bar{a}$ and $\phi$,
\begin{align}
\label{1}
6 \frac{\ddot{\bar{a}}}{\bar{a}} &= -\dot{\phi}^2+V_0 e^{-\lambda} +
\frac{(2-3\gamma) \rho_0}{2 \bar{a}^{3 \gamma}} e^{\sqrt{3}
  (\gamma-4/3) \phi/2}\\
\label{2}
\ddot{\phi} + 3 \frac{\dot{\bar{a}}}{\bar{a}} \dot{\phi} &= \lambda V_0
e^{-\lambda \phi} - \frac{\sqrt{3} (\gamma-4/3) \rho_0}{2 \bar{a}^{3
    \gamma}} e^{\sqrt{3} (\gamma-4/3) \phi/2}
\end{align}
and the constraint equation
\be
\label{3}
6 \frac{\dot{\bar{a}}^2}{\bar{a}^2} = \frac{\dot{\phi}^2}{2}+ V_0
e^{-\lambda \phi} -\frac{\kappa}{\bar{a}^2} + \frac{\rho_0}{\bar{a}^{3
    \gamma}} e^{\sqrt{3} (\gamma-4/3) \phi/2}
\ee
where
\be
\nonumber
V_0 = \frac{\delta}{(1+\delta)}  \ \text{sign}(R)
\qquad \qquad \text{and} \qquad \qquad
\lambda = \frac{(\delta-1)}{\sqrt{3} \delta}.
\ee
Over-dots here denote differentiation with respect to $\eta$.  In
rescaling time in this way we must remember that the more
physically significant cosmological time coordinate $t$ may diverge at
finite values of the new time coordinate.  In such cases we will
consider $t$ as the more physically meaningful time, and allow the new
coordinate to take values only over an appropriate range.  This set of
equations is similar to those obtained in GR, when considering
homogeneous and isotropic cosmologies with a scalar field in an exponential potential, and a perfect fluid.
This similarity is due to the conformal equivalence between $f(R)$ theories of
gravity and GR (see e.g.  \cite{CB, maeda,thesis}).  One
should notice, however, that the terms corresponding to the perfect
fluid have a more complicated form than may have otherwise been
obtained in GR (except for the special case of black-body radiation,
$\gamma=4/3$, which is conformally invariant).  The corresponding set
of equations in GR has been solved in vacuum by Russo
\cite{Russo}, and in the presence of a perfect fluid by Dehnen,
Gavrilov and Melnikov \cite{Russians}.

\section{Vacuum solutions}
\label{vacuum}

\subsection{\textbf{Spatially flat solutions}}

In the absence of any matter fields the equations (\ref{1}) and
(\ref{2}) are identical to homogeneous and isotropic general
relativistic cosmologies, with a scalar field in an exponential
potential.  The spatially flat general solutions to these equations
have already been found by Russo \cite{Russo}.  Here we will briefly
reiterate the method of solving these equations, and use the results
to find the general solution to spatially flat vacuum cosmologies in
$R^n$ gravity.

Firstly, making the transformation of variables $\{a,\phi\}
\rightarrow \{u,v\}$ by the definitions
\be
\nonumber
\bar{a}^3 \equiv e^{v+u} \qquad \qquad \text{and} \qquad \qquad \phi
\equiv \sqrt{\frac{4}{3}} (v-u)
\ee
and defining the new time coordinate
\be
\nonumber
d\eta \equiv e^{\frac{\lambda}{2} \phi} d\tau
\ee
allows the evolution equations (\ref{1}) and (\ref{2}) to be written
as
\begin{align}
\label{1b}
\ddot{v} + \left(1-\frac{\lambda}{\sqrt{3}} \right) \dot{v}^2-\frac{3 V_0}{8} \left(
1+\frac{\lambda}{\sqrt{3}} \right) &= 0\\
\ddot{u} + \left(1+\frac{\lambda}{\sqrt{3}} \right) \dot{u}^2-\frac{3 V_0}{8} \left(
1-\frac{\lambda}{\sqrt{3}} \right) &= 0 \label{2b}
\end{align}
where over-dots now denote differentiation with respect to $\tau$.  In
terms of these new variables, the constraint equation (\ref{3}) reads
\be
\label{3b}
\dot{u} \dot{v} = \frac{3 V_0}{8}.
\ee
We will first solve for two special cases, and then give the solutions
for more general cases.

\subsubsection{\boldmath{$\delta=-1/2$}}

The special case $\lambda=\sqrt{3}$ (corresponding to $\delta=-1/2$) gives the solutions to equations
(\ref{1b}) and (\ref{2b}), under the constraint (\ref{3b}), as
\be
\nonumber
u= c_1 +\frac{1}{2} \ln (\tau+c_2) \qquad \qquad \text{and} \qquad
\qquad v=c_3 +\frac{3 V_0}{8} (\tau+c_2)^2 
\ee
where the $c_i$ are constants.  Transforming these
results back, and absorbing constants into redefinitions of the
coordinates, gives the metric
\be
\label{vac1/2}
\text{
\fbox{\parbox{150pt \fboxsep \fboxrule}{
\be
\nonumber
ds^2 = -\frac{e^{\pm \tau^2}}{\tau^{2/3}} d\tau^2+ \tau^{2/3} d\bf{x}^2
\ee
}}}\ee
where $d\bf{x}^2$ is the line-element of flat three dimensional
Euclidean space.  The $\pm$ sign here is due to the dependence of
$V_0$ on the sign of the Ricci scalar.

\subsubsection{\boldmath{$\delta=1/4$}}

The case $\lambda = -\sqrt{3}$ (corresponding to $\delta=1/4$) is also special and gives
the solutions
\be
\nonumber
u=c_4 +\frac{3 V_0}{8} (\tau+c_5)^2   \qquad \qquad \text{and} \qquad
\qquad 
v= c_6 +\frac{1}{2} \ln (\tau+c_5)
\ee
i.e. the same as the solutions in the previous case, under the
exchange of $u$ and $v$.  Now, transforming variables back to the
originals and absorbing constants into coordinate freedoms, we have
\be
\label{vac1/4}
\text{
\fbox{\parbox{180pt \fboxsep \fboxrule}{
\be
\nonumber
ds^2 = -e^{\pm 2 \tau^2} \tau^{2/3} d\tau^2 + e^{\pm \tau^2} d\bf{x}^2.
\ee
}}}\ee
Here the two $\pm$ signs must be chosen together.

\subsubsection{\boldmath{$\delta \neq -1/2$ or $1/4$}}

Under the conditions $V_0 (\lambda^2-3) >0$ and $\delta \neq -1/2$ or
$1/4$, the solutions to equations (\ref{1b}), (\ref{2b}) and (\ref{3b}) are
\begin{align*}
u &=c_7 +\frac{\sqrt{3}}{\sqrt{3}+\lambda} \ln \left[ \cos \left\{ \sqrt{\frac{3 V_0}{8} \left(
    \frac{\lambda^2}{3}-1\right)} (\tau+c_8) \right\} \right]\\
v &=c_9 +\frac{\sqrt{3}}{\sqrt{3}-\lambda} \ln \left[ \sin \left\{ \sqrt{\frac{3 V_0}{8} \left(
    \frac{\lambda^2}{3}-1\right)} (\tau+c_8) \right\} \right].
\end{align*}
Transforming back to the original variables, and again absorbing constants, gives
\be
\label{vacosc}
\text{
\fbox{\parbox{200pt \fboxsep \fboxrule}{
\be
\nonumber
ds^2= -\frac{\cos^{\frac{2}{4 \delta-1}} \tau}{
\sin^{\frac{2}{2 \delta+1}} \tau} d\tau^2 + \cos^{\frac{4
    \delta}{4\delta-1}}\tau d\bf{x}^2.
\ee
}}}\ee
In the limit $\delta \rightarrow 0$ this solution can be seen to
approach Minkowski space.

Alternatively, when $V_0 (\lambda^2-3) <0$ the solutions are
\begin{align*}
u &= c_{10}+\frac{\sqrt{3}}{\sqrt{3}+\lambda} \ln \left[e^{\sqrt{\frac{3
    V_0}{8} \left(1-\frac{\lambda^2}{3}\right)} \tau}-c_{11} e^{-\sqrt{\frac{3
    V_0}{8} \left(1-\frac{\lambda^2}{3}\right)} \tau} \right]\\
v&= c_{12}+\frac{\sqrt{3}}{\sqrt{3}-\lambda} \ln \left[e^{\sqrt{\frac{3
    V_0}{8} \left(1-\frac{\lambda^2}{3}\right)} \tau}+c_{11} e^{-\sqrt{\frac{3
    V_0}{8} \left(1-\frac{\lambda^2}{3}\right)} \tau} \right]
\end{align*}
which in terms of the original quantities corresponds to the metric
\be
\label{vacpow}
\text{
\fbox{\parbox{290pt \fboxsep \fboxrule}{
\be
\nonumber
ds^2 = -\frac{\left( e^\tau - c_{11} e^{-\tau} \right)^{\frac{2}{4\delta-1}}}
{ \left( e^\tau + c_{11} e^{-\tau} \right)^{\frac{2}{2\delta+1}}} d\tau^2
+ \left( e^\tau - c_{11} e^{-\tau} \right)^{\frac{4\delta}{4\delta-1}} d\bf{x}^2.
\ee
}}}\ee
Here we have absorbed the constants $c_{10}$ and $c_{12}$ into
coordinate redefinitions whilst retaining $c_{11}$.  It should be
noticed that whilst the magnitude of $c_{11}$ could have been absorbed, it has been
left as the special case $c_{11}=0$ corresponds to the attractor
solution.  Again, this solution approaches Minkowski space as $\delta
\rightarrow 0$.

The conditions imposed upon $V_0$ and $\lambda$ here do not imply any
restriction on $\delta$.  This is not immediately obvious as the sign
of $V_0$ depends on the sign of $R$, and hence on the solution.
However, calculating the Ricci scalar for solutions (\ref{vacosc}) and
(\ref{vacpow}), and substituting this into the definition of $V_0$,
shows that both of these solutions exist for any given $\delta$ ($\neq
-1/2$ or $1/4$).

\subsection{\textbf{Spatially curved solutions}}

\subsubsection{\boldmath{$\delta = 1/2$}}

We will now present the general solution for a spatially
curved vacuum universe, when $\delta=1/2$.  When $\rho_0=0$ the change
of variables
\be
\nonumber
\bar{a} \equiv (u v)^{1/4} \qquad \qquad \text{and} \qquad \qquad \phi \equiv
\sqrt{\frac{3}{4}} \ln \left( \frac{u}{v} \right),
\ee
and the new time coordinate $d\tau \equiv \bar{a} d\eta$, allow the field
equations (\ref{1}), (\ref{2}) and (\ref{3}) to be recast in the
simple form
\begin{align*}
\ddot{u} &=0\\
\ddot{v} &= \frac{2}{3} V_0\\
\dot{u} \dot{v} &= \frac{2}{3} \left(V_0 u -\kappa \right).
\end{align*}
Solving these equations yields
\be
\nonumber
u = c_{13} (\tau-c_{14}) \qquad \qquad \text{and} \qquad \qquad v =
\frac{1}{3} (\tau-c_{14})^2 V_0 - \frac{2}{3} \frac{\kappa}{c_{13}}
(\tau-c_{15})
\ee
which on transformation back to the original variables gives
\be
\nonumber
ds^2 = -\frac{d\tau^2}{\tau} + \left\{c_{16} -\kappa \tau
+\frac{V_0}{2} \tau^2 \right\} \left( \frac{dr^2}{1-\kappa r^2}+r^2 d
\Omega^2 \right)
\ee
where $c_{16}=\kappa c_{13} (c_{15}-c_{14})$ and all other constants
have been absorbed into coordinate redefinitions.  In this case we can
transform time coordinates back to proper time $t$, giving
\be
\label{vaccurv}
\text{
\fbox{\parbox{280pt \fboxsep \fboxrule}{
\be
\nonumber
ds^2 = -dt^2 + \left\{c_{17} -\kappa t^2
\pm t^4 \right\} \left( \frac{dr^2}{1-\kappa r^2}+r^2 d
\Omega^2 \right).
\ee
}}}\ee
As well as this solution there exists a second with
\be
\nonumber
u = \frac{\kappa}{V_0} \qquad \qquad \text{and} \qquad \qquad v =
c_{18}+c_{19} \tau + \frac{V_0}{3} \tau^2.
\ee
Transforming this solution back gives Milne space.

\section{Perfect fluid solutions}

Perfect fluid dominated solutions to equations (\ref{1}), (\ref{2}) and
(\ref{3}) will now be presented.  These solutions are for spatially
flat ($\kappa=0$) cosmologies.

\subsection{\boldmath{$\delta=1/(3 \gamma-1)$}}

For the case $\delta=1/(3\gamma -1)$ we can integrate the field
equations exactly, to find the general solution.  Notable exceptions
are given by $\gamma=0$ and $\gamma=1/3$.  For $\gamma=0$, often
associated with vacuum energy, the corresponding value of $\delta$ is
$-1$, which gives a gravitational theory derived from a Lagrangian of
the form $R^0=$constant.  Clearly this is of little interest.  For
$\gamma \rightarrow 1/3$ the value of $\delta \rightarrow \infty$, and
we have a theory that is not clearly defined.  Having excluded these
cases we continue by changing the time coordinate to 
\be
\nonumber
d\tau\equiv d\eta \bar{a}^3 e^{-\lambda \phi}, 
\ee
and considering the variables $u$ and $v$ defined by the transformations
\be
\nonumber
\bar{a} \equiv u^{\frac{1}{2 (3+\sqrt{3}\lambda)}} v^{\frac{1}{2
      (3-\sqrt{3}\lambda)}} \qquad \qquad \text{and} \qquad \qquad
  \phi \equiv \frac{\ln v}{(\sqrt{3}-\lambda)} -
  \frac{\ln u}{(\sqrt{3}+\lambda)}.
\ee
These definitions force us to exclude $\lambda = \pm \sqrt{3}$
(corresponding to $\gamma = -1/3$ and $\gamma = 5/3$).  When $\lambda
= 2/\sqrt{3}-\sqrt{3} \gamma$ (corresponding to $\delta=1/(3\gamma
-1)$)  we can then write the field equations (\ref{1}), (\ref{2}) and
(\ref{3}) as
\begin{align}
\label{u}
\ddot{u} &= -\frac{1}{2} \gamma (5-3 \gamma)
\rho_0 v^{-\frac{1+6\gamma}{1+3\gamma}}\\
\nonumber
\ddot{v} &= 0\\
\nonumber
\dot{u} \dot{v} &= \frac{1}{6} (5-3 \gamma) (1+ 3\gamma) \left( V_0
+\rho_0 v^{-\frac{3 \gamma}{1+3 \gamma}}\right).
\end{align}
Solutions to these equations are
\begin{align*}
u &= \frac{1}{6} (5-3 \gamma)(1+3 \gamma)^2 \rho_0 (\tau-c_{20})^2
(c_{21} (\tau-c_{20}))^{-\frac{1+6\gamma}{1+3 \gamma}} +\frac{(5-3
  \gamma) (1+3 \gamma) V_0}{6 c_{21}} (\tau-c_{22})\\
v &= c_{21} (\tau-c_{20}).
\end{align*}
Transforming back to the original variables, and absorbing constants
into coordinate redefinitions, gives the metric
\be
\label{per1}
\text{
\fbox{\parbox{370pt \fboxsep \fboxrule}{
\begin{align}
\nonumber
ds^2 &= -\tau^{-\frac{6 \gamma}{(1+3\gamma)}} \left((1+3
\gamma) \rho_0 \tau^{\frac{1}{(1+3\gamma)}} +V_0 (\tau+c_{23})
\right)^{-\frac{6 (1-\gamma)}{5-3\gamma}} d\tau^2 \\ \nonumber
& \qquad \qquad \qquad \qquad \qquad + \left((1+3
\gamma) \rho_0 \tau^{\frac{1}{(1+3\gamma)}} +V_0 (\tau+c_{23})
\right)^{\frac{2}{5-3\gamma}} d\bf{x}^2
\end{align}
}}} \ee
where $c_{23} = c_{21} (c_{20}-c_{22})$.  This solution is valid
for all $\gamma \neq -1/3, 0, 1/3$ or $5/3$.

A second solution to the equations (\ref{u}) also exists, when
$V_0<0$.  This solution is given by
\begin{align*}
u &= c_{24}+c_{25} \tau-\frac{1}{4} \gamma (5-3 \gamma) \rho_0 \left(
  -\frac{V_0}{\rho_0}\right)^{\frac{1+6\gamma}{3 \gamma}}  \tau^2\\
v &= \left(-\frac{V_0}{\rho_0}\right)^{-\frac{1+3\gamma}{3\gamma}}.
\end{align*}
Which corresponds to the metric
\be
\label{per2}
\text{
\fbox{\parbox{355pt \fboxsep \fboxrule}{
\begin{align}
\nonumber
ds^2 &= - \left((c_{24} +c_{25} \tau) -\frac{1}{4} \gamma (5-3 \gamma)
  \rho_0 \tau^2 \right)^{-\frac{6 (1-\gamma)}{5-3\gamma}} d\tau^2 \\
\nonumber &\qquad \qquad \qquad \qquad + \left((c_{24} +c_{25} \tau)- \frac{1}{4} \gamma (5-3 \gamma) 
  \rho_0 \tau^2 \right)^{\frac{2}{5-3 \gamma}} d\bf{x}^2.
\end{align}
}}}\ee
Again, this solution is valid
for all $\gamma \neq -1/3, 0, 1/3$ or $5/3$.

\subsection{\boldmath{$\delta = -(4-3\gamma)/(2 (7-3 \gamma))$}}

For the theories $\delta = -(4-3\gamma)/(2 (7-3 \gamma))$ we can
also integrate the field equations directly, to obtain general
solutions for spatially flat cosmologies.  We should note that
the particular cases $\gamma=4/3$, $\gamma=7/3$ and $\gamma=10/3$ are not
usefully solved for here.  Unfortunately, these exceptional cases
include the physically interesting case of $\gamma=4/3$, a fluid of
black-body radiation.  Here $\gamma=4/3$ corresponds to $\delta=0$,
which is GR.  The case $\gamma \rightarrow 7/3$ corresponds to the
limit $\delta \rightarrow \infty$, and $\gamma=10/3$ corresponds to
$\delta =-1$.  Neither of these values of $\delta$ are of physical
interest.

Making the transformation to the variables $u$ and $v$ via the
definitions
\be
\nonumber
a \equiv e^{\left(\frac{\lambda}{\sqrt{3}} u+v\right)} \qquad \qquad \text{and}
  \qquad \qquad \phi \equiv \sqrt{12} \left(
  \frac{\lambda}{\sqrt{3}}v+u \right)
\ee
and redefining time as $d\eta \equiv a^3 d\tau$ allows the field
equations (\ref{1}), (\ref{2}) and (\ref{3}) to be written as
\begin{align*}
\ddot{u} &= \frac{(4-3 \gamma)}{12} \rho_0 e^{\frac{4 (5-3
    \gamma)}{(4-3\gamma)} u}\\
\ddot{v} &= \frac{1}{2} V_0 e^{-\frac{24 (5-3 \gamma)}{(4-3 \gamma)^2}
    v}\\
\dot{u}^2-\dot{v}^2 &= \frac{(4-3\gamma)^2}{24 (5-3 \gamma)} \rho_0
    e^{\frac{4 (5-3 \gamma)}{(4-3 \gamma)} u} + \frac{(4-3
    \gamma)^2}{24 (5-3 \gamma)} V_0 e^{-\frac{24 (5-3 \gamma)}{(4-3 \gamma)^2} v}
\end{align*}
where we have taken
\be
\nonumber
\frac{\lambda}{\sqrt{3}} = -\frac{3 (2-\gamma)}{(4-3\gamma)},
\ee
which corresponds to $\delta = -(4-3\gamma)/(2 (7-3 \gamma))$.  The
case $\gamma=5/3$ does not give a sensible limit, and so we will
exclude it from consideration.  When
$\gamma \neq 5/3$ solutions to these equations are given by
\begin{align*}
u &= - \frac{(4-3\gamma)}{4 (5-3\gamma)} \ln \left[ \frac{(5-3\gamma)
    \rho_0}{6 c_{26}^2} \sin^2\left( c_{26} (\tau-c_{27})\right)
    \right]\\
v &= \frac{(4-3\gamma)^2}{24 (5-3\gamma)} \ln \left[ -\frac{(5-3
    \gamma) V_0}{6 c_{26}^2} \sin^2\left(\frac{6 c_{26}
    }{(4-3\gamma)} (\tau-c_{28})\right) \right].
\end{align*}
The two $\sin$ functions here can be transformed to $\cos$ functions
by suitably redefining the constants $c_{27}$ and $c_{28}$.  These
$\sin$ or $\cos$ functions can then be transformed to $\sinh$ or
$\cosh$ function by the transformation $c_{26} \rightarrow i c_{26}$.
It is, of course, important that solutions remain real for at least
some range of $\tau$.  This restricts which functions should be taken
as physically interesting (assuming $\rho_0>0$), so that the form of the
corresponding metric depends upon the sign of $5-3\gamma$.  We will
treat the different cases separately below.

\subsubsection{\boldmath{$5-3 \gamma >0$}}

The condition $5-3 \gamma >0$ includes the important cases of
pressureless dust and vacuum energy density, and leads to the metric
\be
\label{per3}
\text{
\fbox{\parbox{160pt \fboxsep \fboxrule}{
\be
\nonumber
ds^2 = -b_i^2(\tau) d\tau^2+a_i^2(\tau) d\bf{x}^2
\ee
}}}\ee
where $i=1$, $2$ or $3$.  We then have
\begin{align*}
a_1 &= \sin^{-\frac{1}{(5-3\gamma)}} \{\tau-c_{29}\} \sin^{-\frac{(4-3\gamma)}{6
    (5-3\gamma)}} \left\{ \frac{6}{(4-3\gamma)} \tau\right\}\\
b_1 &= \sin^{-\frac{(7-3\gamma)}{(5-3\gamma)}} \{\tau-c_{29}\} \sin^{\frac{(1-\gamma)(4-3\gamma)}{2
    (5-3\gamma)}} \left\{ \frac{6}{(4-3\gamma)} \tau\right\},
\end{align*}
where constants have been absorbed into coordinate redefinitions, and
$c_{29}=c_{26} (c_{27}-c_{28})$.  These $\sin$ functions can
be transformed to $\cos$ functions through redefinitions of the origin of
the time coordinate $\tau$, and the constant $c_{29}$.  A second
solution also exists with
\begin{align*}
a_2 &= \sinh^{-\frac{1}{(5-3\gamma)}} \{\tau-c_{29}\} \sinh^{-\frac{(4-3\gamma)}{6
    (5-3\gamma)}} \left\{ \frac{6}{(4-3\gamma)} \tau\right\}\\
b_2 &= \sinh^{-\frac{(7-3\gamma)}{(5-3\gamma)}} \{\tau-c_{29}\} \sinh^{\frac{(1-\gamma)(4-3\gamma)}{2
    (5-3\gamma)}} \left\{ \frac{6}{(4-3\gamma)} \tau\right\},
\end{align*}
and a third with
\begin{align*}
a_3 &= \sinh^{-\frac{1}{(5-3\gamma)}} \{\tau-c_{30}\} \cosh^{-\frac{(4-3\gamma)}{6
    (5-3\gamma)}} \left\{ \frac{6}{(4-3\gamma)} \tau\right\}\\
b_3 &= \sinh^{-\frac{(7-3\gamma)}{(5-3\gamma)}} \{\tau-c_{30}\} \cosh^{\frac{(1-\gamma)(4-3\gamma)}{2
    (5-3\gamma)}} \left\{ \frac{6}{(4-3\gamma)} \tau\right\}.
\end{align*}
These solutions are valid for all $\gamma<5/3$ and $\neq 4/3$.

\subsubsection{\boldmath{$5-3\gamma <0$}}

The condition $5-3\gamma <0$ contains the important case of a scalar
field, $\gamma=2$, and gives the metric
\be
\label{per4}
\text{
\fbox{\parbox{160pt \fboxsep \fboxrule}{
\be
\nonumber
ds^2 = -b_j^2(\tau) d\tau^2+a_j^2(\tau) d\bf{x}^2
\ee
}}}\ee
where $j=4$ or $5$.  Here
\begin{align*}
a_4 &= \cosh^{-\frac{1}{(5-3\gamma)}} \{\tau-c_{31}\} \sinh^{-\frac{(4-3\gamma)}{6
    (5-3\gamma)}} \left\{ \frac{6}{(4-3\gamma)} \tau\right\}\\
b_4 &= \cosh^{-\frac{(7-3\gamma)}{(5-3\gamma)}} \{\tau-c_{31}\} \sinh^{\frac{(1-\gamma)(4-3\gamma)}{2
    (5-3\gamma)}} \left\{ \frac{6}{(4-3\gamma)} \tau\right\}
\end{align*}
and a second solution is
\begin{align*}
a_5 &= \cosh^{-\frac{1}{(5-3\gamma)}} \{\tau-c_{32}\} \cosh^{-\frac{(4-3\gamma)}{6
    (5-3\gamma)}} \left\{ \frac{6}{(4-3\gamma)} \tau\right\}\\
b_5 &= \cosh^{-\frac{(7-3\gamma)}{(5-3\gamma)}} \{\tau-c_{32}\} \cosh^{\frac{(1-\gamma)(4-3\gamma)}{2
    (5-3\gamma)}} \left\{ \frac{6}{(4-3\gamma)} \tau\right\}.
\end{align*}
These solutions are valid for all $\gamma>5/3$ and $\neq 7/3$ or $10/3$.

\section{Analysis of vacuum cosmologies}

In this section we will perform an analysis of the solutions found in
section \ref{vacuum}.  The special cases of $\delta = -1/2$ and
$\delta = 1/4$ will be investigated in the appendix.  These solutions
contain a number of constants, as a result of integrating the field
equations.  The physical significance of these constants varies:  In
some cases they can be absorbed into a rescaling of coordinates,
in others they cannot and must be specified by initial
conditions\footnote{This is not a new phenomenon, and can be shown to
  be the case in, for example, the general Friedmann solutions of
  Brans-Dicke theory \cite{BD}.}.  If the latter is the case, then
these constants are physically meaningful quantities and their value
is important for the evolution of the space-time.  We will describe
the effect of taking different values for these constants in the
analysis below.

\subsection{\bf{Spatially flat cosmologies}}

We begin with the spatially flat solution (\ref{vacpow}).  In the case $c_{11}=0$
the solution (\ref{vacpow}) can be written as
\begin{align*}
ds^2 &= -e^{-\frac{4 (\delta-1)}{(4 \delta-1) (2 \delta+1)} \tau}
d\tau^2 + e^{\frac{4 \delta}{(4 \delta-1)}\tau} d\bf{x}^2\\
&= -dt^2+t^{\frac{2 \delta (1+2\delta)}{(1-\delta)}} d\bf{x}^2,
\end{align*}
which is simply the power-law solution (\ref{power2}).  It can be seen
directly from (\ref{vacpow}) that for all $c_{11}$ this simple
solution is the attractor as $\tau \rightarrow \infty$.

When $c_{11}=0$ the power-law solution above describes the evolution
of the universe all the way back to the initial singularity, but for
$c_{11} \neq 0$ more general behaviour occurs as the singularity is
approached.  The form of this early-time behaviour depends upon the
sign of $c_{11}$ (recall that the magnitude of $c_{11}$ can be
absorbed into coordinate redefinitions).

For $c_{11} >0$ a power series expansion of the metric gives
\begin{align*}
ds^2 &\simeq -\tau^{\frac{2}{(4\delta-1)}} d\tau^2 +\tau^{\frac{4
    \delta}{(4\delta-1)}} d\bf{x}^2\\
&= -dt^2 + t d\bf{x}^2
\end{align*}
when $\tau -\frac{1}{2} \log (c_{11})\ll 1$.  This limit is the same as the power-law exact
solution (\ref{power3}).

For $c_{11} <0$ a similar expansion about $\frac{1}{2} \log (-c_{11})$ gives
\begin{align*}
ds^2 &\simeq -\tau^{\frac{-2}{(1+2\delta)}} d\tau^2 +\left(
1+\frac{\delta}{(4 \delta-1)} \tau^2 \right)^2 d\bf{x}^2\\
&= -dt^2 +\left(1+\frac{\delta}{(4 \delta-1)} t^{\frac{(1+2
    \delta)}{\delta}}  \right)^2 d\bf{x}^2.
\end{align*}
This solution is clearly quite different from the power-law limits that
have been found so far.  For $\delta > 1/4$ or $<0$ it corresponds
to a non-zero minimum of expansion, or bounce.  For $0< \delta <1/4$
it is a maximum of expansion.  It should be noted that when
$-1/2 < \delta <0$ the power of $t$ in the scale factor diverges as
$t\rightarrow 0$.  In this case there is no staticity, and the
scale-factor diverges as $t \rightarrow 0$.  Bounces occur here
without any violation of energy conditions, as
there are no matter fields present to commit such violations.  The
bounce is entirely due to the vacuum dynamics of the theory.

\begin{figure}[ht]
\center \epsfig{file=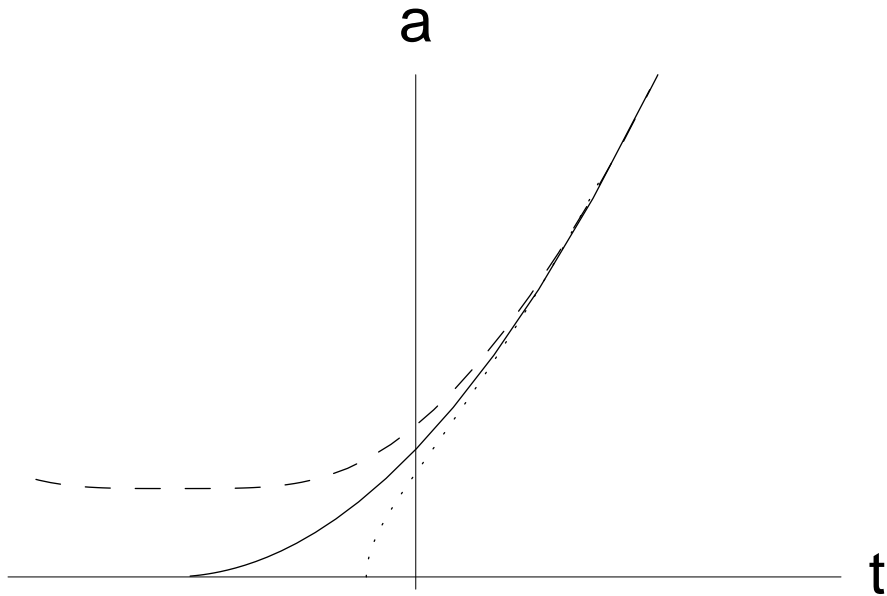,height=8cm}
\caption{The evolution of the scale factor of solution
  (\ref{vacpow}) in terms of proper time, $t$.  The value of $\delta$
  chosen here is $1/2$.  The solid line corresponds to $c_{11}=0$, or the power-law
  exact solution (\ref{power2}).   The dotted line corresponds to
  $c_{11}>0$ and shows an early period of $t^{1/2}$ expansion, that is
  attracted towards (\ref{power2}) at later times.  The dashed line corresponds
  to $c_{11}<0$ and shows a non-zero minimum of expansion, and
  subsequent evolution towards (\ref{power2}).}
\label{vacpowplot}
\end{figure}

We have now shown both the early and late-time behaviour of the
solution (\ref{vacpow}).  At late-times all solutions approach the
power-law exact particular solution (\ref{power2}).  At early-times
the evolution of the universe is qualitatively different depending
upon the value of the constant $c_{11}$.  For $c_{11}=0$ the solution
(\ref{power2}) is valid all the way back to the initial singularity.
For $c_{11}>0$ the general solution has an early period of expansion of
the form of (\ref{power3}), in the vicinity of the singularity.  The
expansion then evolves into the late-time attractor.  When $c_{11}<0$
there is a non-zero minimum of expansion (or maximum if
$0<\delta<1/4$, or divergence if $-1/2<\delta<0$).  The universe then evolves from this static past
towards the late-time attractor.  Some representative evolutions of the
scale-factor, in terms of the proper time coordinate $t$, are shown in
figure \ref{vacpowplot}.

We will now investigate the solution (\ref{vacosc}).  Firstly,
expanding around the point $\tau=0$ we get the approximation
\begin{align*}
ds^2 &\simeq -\tau^{-\frac{1}{(1+2\delta)}} d\tau^2 + \left(1-
\frac{4\delta}{(4\delta-1)} \tau^2 \right)^2 d\bf{x}^2\\
&= -dt^2 + \left(1- \frac{4\delta}{(4\delta-1)}
t^{\frac{(1+2\delta)}{\delta}} \right)^2 d\bf{x}^2,
\end{align*}
when $\tau \ll 1$.  This is remarkably similar to the form of the
previous metric, (\ref{vacpow}), in the vicinity of its minimum of
expansion.  A noticeable difference is the sign before the second term
in the brackets.  This change in sign shows that when the previous
solution was a minimum ($\delta >1/4$ or $<-1/2$) this solution is a
maximum of expansion.  Correspondingly, when the previous solution was
a maximum ($0<\delta<1/4$), this solution is a minimum.  Again, there
is divergence when $-1/2<\delta<0$.

We will now investigate the form of (\ref{vacosc}) around $\tau=\pm
\pi/2$.  Performing a power series expansion about either of these
points gives
\begin{align*}
ds^2 &\simeq -\tau^{\frac{2}{(4\delta-1)}} d\tau^2 -
\tau^{\frac{4\delta}{(4\delta-1)}} d\bf{x}^2\\
&= -dt^2 +t d\bf{x}^2,
\end{align*}
when $\tau\mp\pi/2 \ll 1$.  In the vicinity of both of these points the
evolution of (\ref{vacpow}) is therefore of the form of the
power-law exact solution (\ref{power3}).

\begin{figure}[ht]
\center \epsfig{file=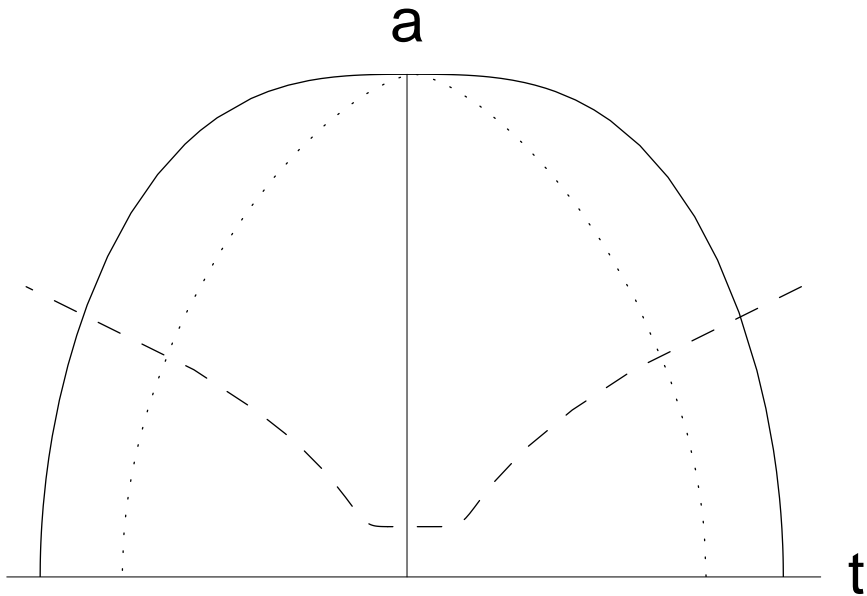,height=8cm}
\caption{The evolution of the scale factor of solution
  (\ref{vacosc}) in terms of proper time, $t$.  The solid line
  corresponds to $\delta=1$, the dashed line to $\delta=1/8$ and the
  dotted line to $\delta=-2$.}
\label{vacoscplot}
\end{figure}

We have shown that the evolution of the solution (\ref{vacosc})
depends upon the sign of $\delta$.  When $\delta>1/4$ or $<-1/2$ the
solution, in terms of proper time $t$, has a maximum of expansion with
its early and late-time evolution going as $t^{\frac{1}{2}}$.  When
$0<\delta<1/4$ there is a minimum of expansion, and late and
early-time expansion again goes as $t^{1/2}$.  For $-1/2<\delta<0$ the
solution diverges in the region which is static for all other values
of $\delta$.  The form of some representative solutions are shown in
figure \ref{vacoscplot}.  The behaviour of these solutions are markedly different from
those found by Russo \cite{Russo}.  Although a late-time power-law
attractor can still be seen to exist in one of the solutions, very
little else is comparable.

\subsection{\bf{Spatially curved cosmologies}}

For spatially curved vacuum universes we found the general solutions
(\ref{vaccurv}) and Milne space, when $\delta=1/2$.  The existence of Milne space as a
solution should not be surprising, as any Ricci flat solution of GR in
a vacuum is also a solution of the field equations (\ref{field}) in vacuum (at least
when $\delta>0$).  We will now investigate the form of the solution
(\ref{vaccurv}).

When $\kappa=0$ this solution reduces to
\be
\nonumber
ds^2 = -dt^2 + t^4 d\bf{x}^2,
\ee
which corresponds to the power-law particular solution (\ref{power2}),
when $\delta=1/2$.  When $\kappa \neq 0$ we see that this solution
exhibits new behaviour, dependent on the value of $\kappa$ and which
branch of the $\pm$ sign is chosen.

When $\kappa >0$ and $c_{17}>0$, then in the vicinity of
$t=0$ this solution exhibits a maximum of expansion.  Subsequent
evolution of the scale-factor, as $t\rightarrow \pm \infty$, depends
upon the chosen branch.  Taking the positive branch we see that the
universe can start to expand again, as $\vert t \vert$ increases.
The existence of this expansion phase depends upon the magnitude of
$c_{17}$.  If $c_{17}\leqslant \kappa^2/4$ then the effect of the
second term in the scale-factor is too great and the universe
collapses to a singularity, before the $t^4$ term can become dominant.
When $c_{17} > \kappa^2/4$ the $t^4$ term
dominates the late-time evolution and the power-law solution
(\ref{power2}) is approached, as $\vert t\vert \rightarrow \infty$.
When the negative branch is taken then the maximum of expansion at
$t=0$ always leads to collapse to a singularity, as $t$ either
increases or decreases.

When $\kappa <0$ and $c_{17}=0$ then there exists an initial singularity, with the evolution of the scale
factor in its immediate vicinity expanding proportionally to $t$.  The
late-time evolution of such solutions will be attracted to either the power-law
solution (\ref{power2}), if the positive branch is chosen, or to
eventual collapse to singularity, if the negative branch is chosen.
The case $c_{17} >0$ does not have a singularity at $t=0$, but instead
has a minimum of expansion, or bounce.  The evolution away from this
minimum proceeds as in the $c_{17}=0$ case.  The $c_{17} <0$ case
again features an initial singularity and displays approximately the
same evolution as when $c_{17}=0$ (a notable deviation is that the
period of $a\propto t$ expansion near the singularity is absent if
$c_{17}$ is sufficiently negative).

\begin{figure}[ht]
\center \epsfig{file=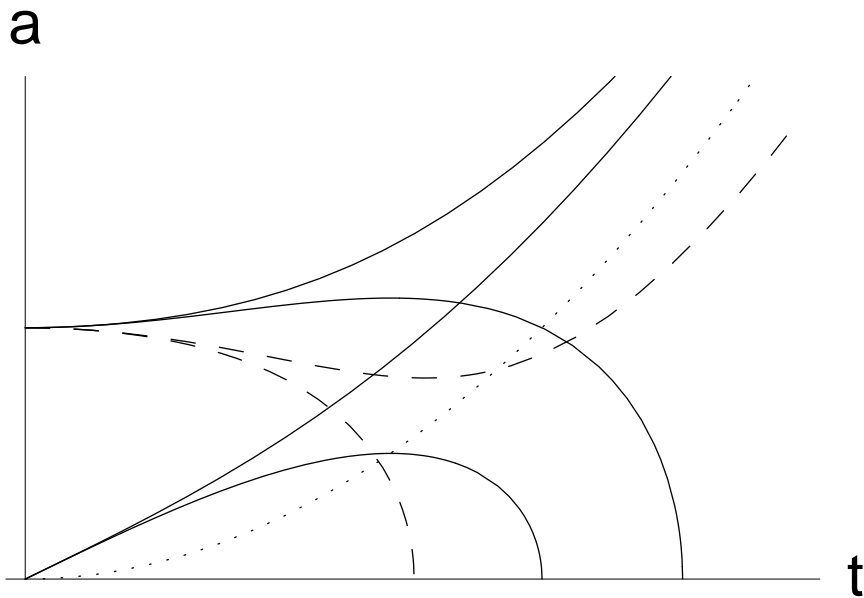,height=8cm}
\caption{The evolution of the scale factor of solution
  (\ref{vaccurv}) in terms of proper time, $t$.  The solid lines
  correspond to universes with $\kappa>0$, dashed lines to those with
  $\kappa<0$ and the dotted line to $\kappa=0$.  Those solutions
  starting from the origin have $c_{17}=0$, and those starting from a
  non-zero minimum have $c_{17}>0$.  All solutions collapsing to
  singularity correspond to the negative branch of $t^4$, and all
  solutions expanding at late-times to the positive branch.  See the
  main body of text for an explanation of these different behaviours.}
\label{vaccurvfig}
\end{figure}

We have shown that when $\kappa \neq 0$ there exists a minimum of
expansion when $c_{17}>0$, and an initial singularity when $c_{17}
\leqslant 0$.  Initial evolution away from an early minimum proceeds as
$a\propto t$, either expanding or collapsing, depending on the sign of
$\kappa$.    If there is an initial singularity, then evolution away
from it proceeds as $a\propto t$, unless $c_{17}$ is sufficiently
negative  (in which case the $t^4$ term dominates right from the
beginning).  The late-time evolution of these solutions is
dominated by the $t^4$ term in (\ref{vaccurv}).  The negative branch
of this term causes collapse to a singularity, and the positive branch
leads to asymptotic expansion and approach towards the
power-law solution (\ref{power2}) (unless the term proportional to
$\kappa$ causes collapse to singularity before the $t^4$ becomes
dominant).  The form of some representative solutions are shown in
figure \ref{vaccurvfig}.

\section{Analysis of perfect fluid cosmologies}

We will now analyse the evolution of universes described by the
perfect fluid solutions (\ref{per1}), (\ref{per2}), (\ref{per3}) and
(\ref{per4}).  Again, these solutions contain a number of constants,
some of which can be absorbed into coordinate redefinitions and some
of which cannot.  We will explain the significance of such constants below.

\subsection{\boldmath{$\delta=1/(3 \gamma-1)$}}

When $\delta=1/(3 \gamma-1)$ we found from direct integration of the
field equations the general solutions (\ref{per1}) and (\ref{per2}).
We will first consider (\ref{per1}).  

The form of the scale factor in this solution depends upon which of
the two terms in the brackets is dominant.  When the first term dominates the
metric is approximately given by
\begin{align*}
ds^2 &\simeq -\tau^{-\frac{6 (1+4\gamma-3\gamma^2)}{(5-3\gamma)
    (1+3\gamma)}} d\tau^2 +\tau^{\frac{2}{(5-3\gamma)(1+3\gamma)}}
d\bf{x}^2\\
&= -dt^2 + t d\bf{x}^2,
\end{align*}
which corresponds to the power-law particular solution
(\ref{power3}).  When the second term in the brackets dominates we have
\begin{align*}
ds^2 &\simeq -\tau^{-\frac{6 (1+7\gamma-6\gamma^2)}{(5-3\gamma)
    (1+3\gamma)}}+\tau^{\frac{2}{(5-3\gamma) (1+3\gamma)}} d\bf{x}^2\\
&= -dt^2 +t^{\frac{2 (1+3 \gamma)}{(2-3\gamma) (1-3 \gamma)}}
    d\bf{x}^2,
\end{align*}
which corresponds to the vacuum dominated, power-law particular
solution (\ref{power2}).  It now remains to investigate the conditions
under which each of these approximations dominates, at early and late times.

When $c_{23}=0$, and $\gamma>0$, then the first term always dominates
as $\tau \rightarrow 0$, and the universe evolves like $a\sim
t^{\frac{1}{2}}$.  As $\tau \rightarrow \infty$ the second term
dominates, and the evolution of the universe is attracted towards the
form of (\ref{power2}).

When $c_{23} \neq 0$, and $\gamma >0$, then this has the effect of
offsetting the origin of the second term relative to the first.  The
second term will then be the dominant one for most of the range of
$\tau$, with a contribution expected from the first term in the
vicinity of $\tau=-c_{23}$.

Phantom fluids, with $\gamma<0$, have the same behaviour as outlined
above, but with the dominance of the two terms reversed in each case.

We will now consider the solution (\ref{per2}).  The evolution of this
space-time again depends on which of the two terms in the brackets is
dominant.  If the first term dominates then the metric look like
\begin{align*}
ds^2 &\simeq -\tau^{-\frac{6 (1-\gamma)}{(5-3\gamma)}} d\tau^2
+\tau^{\frac{2}{(5-3\gamma)}} d\bf{x}^2\\
&= -dt^2+t d\bf{x}^2,
\end{align*}
which is again the power-law evolution (\ref{power3}).  We will call
this vacuum domination, as it corresponds to evolution being dominated
by a term independent of $\rho_0$.  If the second term in brackets
dominates then we now have
\begin{align*}
ds^2 &\simeq -\tau^{-\frac{12 (1-\gamma)}{(5-3\gamma)}} d\tau^2
+\tau^{\frac{4}{(5-3\gamma)}} d\bf{x}^2\\
&= -dt^2 + t^{\frac{4}{(3\gamma-1)}} d\bf{x}^2,
\end{align*}
which corresponds to the matter dominated power-law expansion
(\ref{power}).  We must now investigate the conditions
under which each of these approximations dominates.  This will depend
upon the value of the constants $c_{24}$ and $c_{25}$.

When $c_{24}=c_{25}=0$ we see that the matter dominated power-law
expansion (\ref{power}) holds all the way back to the initial
singularity.  If $c_{24}=0$ and $c_{25} \neq 0$, then there is an
initial period of vacuum dominated expansion of the form $a\propto
t^{\frac{1}{2}}$ and a late-time evolution towards the matter
dominated power-law solution (\ref{power}).  If $c_{24} \neq 0$ then
this has the effect of offsetting the origin of first term in the
brackets, relative to the second.  This can result in a non-zero
minimum of expansion (depending on the signs of $c_{24}$ and $c_{25}$),
as in this case the first term can both dominate and be non-zero at
$\tau=0$.  As $\vert \tau \vert \rightarrow \infty$ the matter
dominated power-law solution (\ref{power}) is then approached.

\subsection{\boldmath{$\delta = -(4-3\gamma)/(2 (7-3 \gamma))$}}

We will now investigate the solutions (\ref{per3}) and (\ref{per4}),
obtained by integrating the field equations in the presence of a
perfect fluid when $\delta = -(4-3\gamma)/(2 (7-3 \gamma))$.

\subsubsection{\boldmath{$5-3\gamma >0$}}

We will first consider the solution $\{ a_1, b_1 \}$.  Expanding this solution around the points
$\tau=n \pi$ and $\tau=c_{29}+n \pi$, where $n$ is an integer, gives the evolution of the scale factor
in their vicinity.  To first order, an expansion about $\tau=n \pi$ results in
\begin{align*}
ds^2 &\simeq -\tau^{\frac{(4-3 \gamma) (1-\gamma)}{(5-3\gamma)}}
d\tau^2 + \tau^{-\frac{(4-3\gamma)}{(5-3\gamma)}} d\bf{x}^2\\
&= -dt^2 + t^{\frac{2 (4-3\gamma)}{3 (7-3 \gamma) (\gamma-2)}} d\bf{x}^2,
\end{align*}
which corresponds to the vacuum dominated power-law expansion,
(\ref{power2}).  A similar expansion about $\tau=c_{29}+n\pi$ gives
\begin{align*}
ds^2 &\simeq -(\tau-c_{29})^{-\frac{2 (7-3\gamma)}{(5-3 \gamma)}}
d\tau^2 + (\tau-c_{29})^{-\frac{2}{(5-3 \gamma)}} d\bf{x}^2\\
&= -dt^2 +td\bf{x}^2,
\end{align*}
which is the power-law evolution described by (\ref{power3}).  Whether
or not these points correspond to a curvature singularity depends upon
the value of $\gamma$, and can be read off from the Ricci scalar
\be
\nonumber
R = \frac{6 (10-3 \gamma)}{(5-3 \gamma) (4-3\gamma)} \sin^{\frac{2
    (7-3 \gamma)}{(5-3\gamma)}} \{\tau-c_{29}\} \sin^{-\frac{(2-\gamma)
    (7-3 \gamma)}{(5-3\gamma)}} \left\{\frac{6}{(4-3\gamma)}
\tau \right\}.
\ee
We see that when $\gamma \geqslant 7/3$ the Ricci scalar remains finite for all
$\tau$, whilst for $2<\gamma<7/3$ or $\gamma<5/3$ the Ricci scalar
diverges to infinity every time $\tau = (4-3 \gamma) n \pi/6$.  The
evolution of $a(t)$ in the vicinity of these
singularities goes like (\ref{power2}).  Similarly, when
$5/3<\gamma<7/3$ there are curvature singularities at $\tau=c_{29} +n
\pi$.  In the vicinity of these singularities $a(t) \sim
t^{\frac{1}{2}}$.  These latter singularities correspond to
$a(t) \rightarrow 0$, whereas the former correspond to the divergence
$a(t) \rightarrow \infty$ (except in the range $4/3<\gamma<5/3$, in
which case $a(t) \rightarrow 0$).

An instructive special case to consider is that of pressureless dust,
$\gamma=0$.  In this case the coordinate transformation $t=\cot
(\tau-c_{29})$ allows the metric to be recast as
\be
\nonumber
ds^2 = -dt^2 +\frac{\sqrt{t^2+1}}{\sin^\frac{1}{6} \{6 (c_{29}
  +\cot^{-1}\{t\})\}} d\bf{x}^2.
\ee
It can now be seen directly that $a(t) \rightarrow t^\frac{1}{2}$ as
$t\rightarrow \infty$ (or $\tau\rightarrow c_{29}+n\pi$), and diverges to
$\infty$ as (\ref{power2}) when $t \rightarrow -\cot \{c_{29}-n\pi/6\}$ (or
$\tau \rightarrow n\pi/6$).

We will now consider the solution $\{ a_2, b_2 \}$.  In the limit
$\tau \rightarrow \pm \infty$ this solution approaches
\begin{align*}
ds^2 &\simeq -e^{-\frac{8 \tau}{(5-3\gamma)}}d\tau^2 +
e^{-\frac{4\tau}{(5-3\gamma)}} d\bf{x}^2\\
&= -dt^2 +t d\bf{x}^2,
\end{align*}
which is the power-law solution (\ref{power3}).  When $\tau\rightarrow
0$ we see that we have the same limit as in the previous solution,
where the power-law evolution (\ref{power2}) is approached.
Similarly, as $\tau\rightarrow c_{29}$ we approach the
power-law evolution $a\sim t^{\frac{1}{2}}$, (\ref{power3}).  The
conditions for whether or not these limits
correspond to singularities, and if these singularities correspond to
$a(t) \rightarrow 0$ or $a(t)\rightarrow \infty$, have exactly the
same dependence on $\gamma$ as with the previous solution.

The pressureless dust case is again instructive, and can be written in
terms of proper time, by making the transformation
$t=\coth\{c_{29}-\tau\}$, as
\be
\nonumber
ds^2 = -dt^2 +\frac{\sqrt{t^2-1}}{\sinh^\frac{1}{6} \{6 (c_{29}
  -\coth^{-1}\{t\})\}} d\bf{x}^2.
\ee
This solution displays $t^\frac{1}{2}$ evolution as $t\rightarrow
\infty$, and divergence to $\infty$ as $t \rightarrow \coth\{c_{29}\}$.

The solution $\{ a_3, b_3 \}$ will now be considered.  The late-time
evolution of this solution is the same as in the previous case.  The
evolution about $\tau=c$ is also the same.  Now, however, we no longer
have the singular behaviour about $\tau=0$ that previously existed.
Instead the scale-factor evolves as $a\sim a_0+t^\frac{1}{2}$ in the
vicinity of this point, where $a_0$ is some constant.  The dust
solution in this case is then
\be
\nonumber
ds^2 = -dt^2 +\frac{\sqrt{t^2-1}}{\cosh^\frac{1}{6} \{6 (c_{30}
  -\coth^{-1}\{t\})\}} d\bf{x}^2,
\ee
where $t=\coth\{c_{30}-\tau\}$.  This metric has a scale factor that
no longer diverges at any finite $t$.

\subsubsection{\boldmath{$5-3\gamma <0$}}

The late-time evolution of these two solutions is of the form
$a\sim t^\frac{1}{2}$, as with the previous two solutions.  Now, the
form of solution \{$a_4$, $b_4$\} about $\tau=0$ displays the same
divergences as \{$a_2$, $b_2$\}.  Evolution towards the point $\tau=c$
is non-singular, and goes as $a\sim a_0+t^\frac{1}{2}$.  The
special case of pressureless dust is given in this case, in terms of
proper time $t$, by
\be
\nonumber
ds^2 = -dt^2 +\frac{\sqrt{1-t^2}}{\sinh^\frac{1}{6} \{6 (c_{31}
  +\tanh^{-1}\{t\})\}} d\bf{x}^2,
\ee
where $t=\tanh\{\tau-c_{31}\}$.  The scale-factor here can be seen to
vanish at $t=\pm1$ (corresponding to $\tau \rightarrow \pm \infty$),
and to diverge to $\infty$ at $t=\tanh\{-c_{31}\}$.  The evolution of
the scale factor follows the late-time attractor towards the points at
$t=\pm 1$.

We will now consider the remaining solution, \{$a_5$, $b_5$\}.  This
solution behaves as the solution \{$a_3$, $b_3$\} in the vicinity of
$\tau=0$, and as \{$a_4$, $b_4$\} in the vicinity of $\tau=c$.  That
is, about both of these points the evolution of the scale factor goes
as $a\sim a_0+t^\frac{1}{2}$ and is non-singular.  In fact, at no
finite $\tau$ does this solution become singular.  The pressureless
dust solution is given here, in terms of proper time, as
\be
\nonumber
ds^2 = -dt^2 +\frac{\sqrt{1-t^2}}{\cosh^\frac{1}{6} \{6 (c_{32}
  +\tanh^{-1}\{t\})\}} d\bf{x}^2,
\ee
where $t=\tanh\{\tau-c_{32}\}$.  This solution describes a space-time
that evolves from $a=0$ at $t=-1$ (corresponding to $\tau = -\infty$)
to $a=0$ at $t=1$ (corresponding to $\tau=\infty$).  The evolution
between these two end points is at all times finite.

\section{Discussion}

We have investigated the homogeneous and isotropic cosmological
solutions of $R^n$ theories.  It has been shown that in a number of
cases the field equations can be integrated directly, allowing the
general behaviour of these models to be found.  For spatially flat
vacuum universes the solutions for any $n$ can be found.  For
spatially curved vacuum universes, and flat perfect fluid universes, the general
solutions can be obtained for various particular values of $n$.  These
solutions were given explicitly in sections 3 and 4.  It is
interesting to note that unlike the general relativistic cosmologies,
vacuum solutions can be either open, closed or flat.  It was also
found that there is generically not one unique solution for any given
$n$, matter content and topology, as there usually is in GR.  This was found previously in
\cite{power,Car04}, where a phase plane analysis showed an invariant
sub-manifold in the phase space of solutions which could not be crossed
by any trajectory.  This result is reiterated here, where we find
multiple ways to integrate the field equations, yielding multiple
solutions.  It was also shown in \cite{power,Car04} that the form of
the general solution changes with $n$:  Critical points in the phase
space change their attractor nature as $n$ is varied.  Again, we confirm this result
by finding explicit solutions for cosmologies with the same topology and
matter content, but different $n$.  These solutions have clearly
different forms.  In sections 5 and 6 we performed a brief analysis of the vacuum and
perfect fluid solutions, respectively.  The evolution of these
universes, in terms of proper time, were found in different regimes.
The nature of the initial singularity was investigated, or its lack thereof, as well as evolution at
late-times.  In summary, it was found that great variety is present in the
cosmologies of these theories.  Behaviour at both late and early times
can vary wildly, depending on $n$, and even between different
solutions with the same $n$, matter content and topology.  Solutions
can manifest initially singular or non-singular behaviour, and
late-time evolution can either lead to a crunch or to monotonic
expansion.

\newpage
\leftline{\bf Acknowledgements}

I would like to thank John Barrow for helpful discussions and suggestions, and
to acknowledge the support of the Lindemann trust.

\appendix
\section{Vacuum \boldmath{$\delta=-1/2$} cosmologies}

\begin{figure}[ht]
\center \epsfig{file=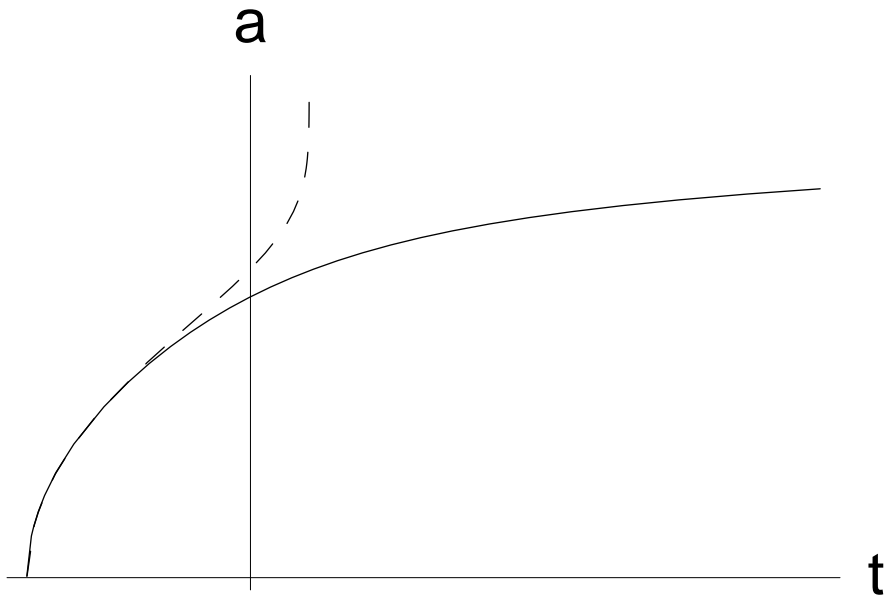,height=8cm}
\caption{The evolution of the scale factor of solution
  (\ref{vac1/2}) in terms of proper time, $t$.  The solid lines
  correspond to the positive branch, and the dashed line to the
  negative branch.}
\label{vac1/2fig}
\end{figure}

We will now describe the evolution of universes described by the
solution (\ref{vac1/2}).  Taking the positive branch of (\ref{vac1/2})
we see that as $\tau \rightarrow \infty$ we have $d\tau \ll dt$,
so that in terms of proper time the universe evolves very slowly at
late-times.  However, for the negative branch of (\ref{vac1/2}) we see that
$d\tau \gg dt$ in this limit.  For the negative branch, therefore, we should
expect increasingly rapid evolution at late times, and for the positive branch
an evolution towards an asymptotic steady state.  A power series
expansion of (\ref{vac1/2}) shows that the early-time behaviour of
both branches goes like the power-law particular solution
(\ref{power3}).  Figure \ref{vac1/2fig} shows graphically the
evolution of the two branches of this solution.

\section{Vacuum \boldmath{$\delta=1/4$} cosmologies}

\begin{figure}[ht]
\center \epsfig{file=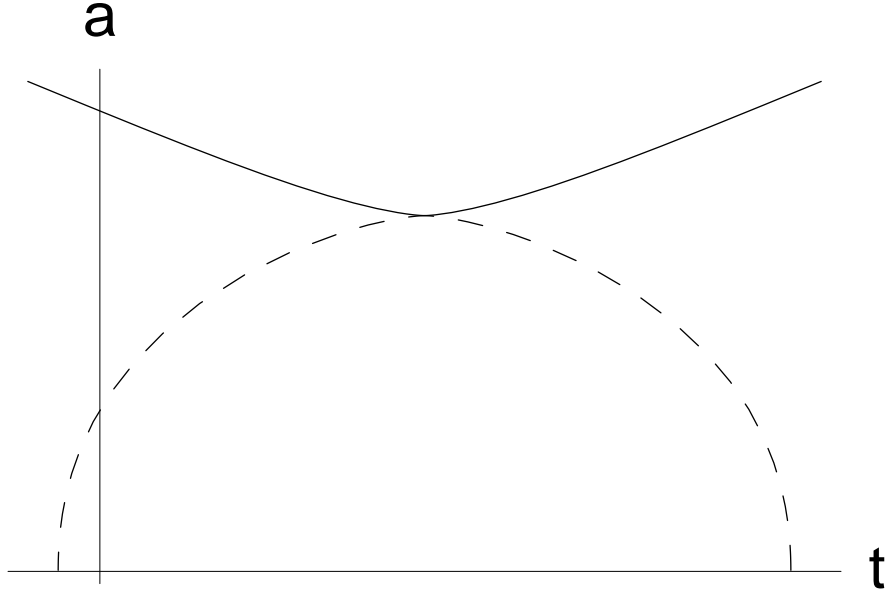,height=8cm}
\caption{The evolution of the scale factor of solution
  (\ref{vac1/4}) in terms of proper time, $t$.  The solid lines
  correspond to the positive branch, and the dashed line to the
  negative branch.}
\label{vac1/4fig}
\end{figure}

We will now describe the evolution of universes described by the
solution (\ref{vac1/4}).  A power series expansion of this solution
shows that the scale factors of both branches approach a
stationary state, in terms proper time $t$, as $\tau \rightarrow 0$.
We can then see that as $\tau$ increases the positive branch expands out of
this stationary state, and the negative branch collapses out of it.
The positive branch therefore corresponds to an expanding universe
evolving out of a bounce, and the negative universe to an initially
expanding universe with a maximum of expansion and subsequent
collapse.  Figure \ref{vac1/4fig} shows the evolution of the two branches of this solution.


\begin{thebibliography}{99}

\bibitem{buch} H. Buchdahl, J. Phys. A \textbf{12}, 1229 (1979).

\bibitem{kerner} R. Kerner, Gen. Rel. Grav. \textbf{14}, 453 (1982).

\bibitem{BO} J. D. Barrow and A. C. Ottewill, J. Phys. A \textbf{16}, 2757
(1983).

\bibitem{magg} G. Magnano, M. Ferraris and M. Francaviglia, Gen. Rel. Grav. 
\textbf{19}, 465 (1987).

\bibitem{pech} E. Pechlaner and R. Sexl, Comm. Math. Phys. \textbf{2},
  165 (1966).

\bibitem{str1} M. Gasperini, M. Maggiore and G. Veneziano,
  Nucl. Phys. B \textbf{494}, 315 (1997).

\bibitem{str2} K. A. Meissner, Phys. Lett. B \textbf{392}, 298 (1997).

\bibitem{tur} S. M. Carroll, A. De Felice, V. Duvvuri, D. A. Easson, M.
Trodden, and M. S. Turner, Phys. Rev. D \textbf{71}, 063513 (2005).

\bibitem{new} S. Nojiri and S. D. Odintsov, Phys. Rev. D \textbf{68},
  123512 (2003).

\bibitem{berk} A. Berkin, Phys. Rev. D \textbf{44}, 1020 (1991).

\bibitem{brun} E. Bruning, D. Coule and C. Xu,
  Gen. Rel. Grav. \textbf{26}, 1197 (1994).

\bibitem{herv} J. D. Barrow and S. Hervik, Phys. Rev. D \textbf{73},
  023007 (2006).  J. D. Barrow and S. Hervik, Phys. Rev. D
  \textbf{74}, 124017 (2006).

\bibitem{brrw} J. D. Barrow and J. Middleton, gr-qc/0702098.

\bibitem{clftn} J. D. Barrow and T. Clifton,
  Class. Quant. Grav. \textbf{23}, L1 (2006).  T. Clifton and
  J. D. Barrow, Class. Quant. Grav. \textbf{23}, 2951 (2006).

\bibitem{dnsby} J. A. Leach, S. Carloni and P. K. S. Dunsby,
  Class. Quant. Grav. \textbf{23}, 4915 (2006).

\bibitem{birk} T. Clifton, Class. Quant. Grav. \textbf{23}, 7445 (2006).

\bibitem{power} T. Clifton and J. D. Barrow, Phys. Rev. D \textbf{72},
  103005 (2005).

\bibitem{godel} T. Clifton and J. D. Barrow, Phys. Rev. D \textbf{72},
  123003 (2005).

\bibitem{Buc70} H. A. Buchdahl, Mon. Not. R. Astron. Soc. \textbf{150}, 1
(1970).

\bibitem{schmidt} H. J. Schmidt, gr-qc/0407095.

\bibitem{Car04} S. Carloni, P. K. S. Dunsby, S. Capoziello and A. Troisi,
Class. Quant. Grav. \textbf{22} (2005) 4839.

\bibitem{Od1}   S. Nojiri and S. D. Odintsov, Phys. Rev. D
  \textbf{74}, 086005 (2006).

\bibitem{Od2}  S. Nojiri and S. D. Odintsov, hep-th/0611071.

\bibitem{ST}  D. Wands, Class. Quant. Grav. \textbf{11}, 269 (1994).

\bibitem{CB} J. D. Barrow and S. Cotsakis, Phys. Lett. B \textbf{214} 515
(1994).

\bibitem{maeda} K-I. Maeda, Phys. Rev. D \textbf{39}, 3159 (1989).

\bibitem{thesis} T. Clifton, PhD thesis, gr-qc/0610071.

\bibitem{Russo} J. G. Russo, Phys. Lett. B \textbf{600}, 185 (2004).

\bibitem{Russians} H. Dehnen, V. R. Gavrilov and V. N. Melnikov,
  Grav. Cosmol. \textbf{9}, 189 (2003).

\bibitem{BD}  L. E. Gurevich, A. M. Finkelstein and V. A. Ruban,
  Astrophys. Space Sci. \textbf{22}, 231 (1973).

\end{thebibliography}
\end{document}